\documentclass{jps-cp}
\usepackage{txfonts} 
\usepackage{siunitx}
\usepackage{color}

\title{Few-Ion Schottky Power Scaling and Automated Reconstruction of Single-Ion Events} 

\author{Moritz \textsc{Porstendörfer}$^{1}$, David \textsc{Freire-Fernández}$^{2,3,1}$, Anna Jule \textsc{Heß}$^{1}$, Wolfram \textsc{Korten}$^{2,3,4}$,  and Yury A. \textsc{Litvinov}$^{1,3}$}

\inst{$^{1}$Institute for Nuclear Physics, University of Cologne, Cologne, Germany \\
$^{2}$FAIR Facility for Antiproton and Ion Research in Europe GmbH, 64291 Darmstadt, Germany
\\
$^{3}$GSI Helmholtzzentrum für Schwerionenforschung GmbH,
64291 Darmstadt, Germany\\
$^{4}$IRFU, CEA, Universit{\'e} Paris-Saclay, 91191 Gif-sur-Yvette, France}

\email{D.FreireFernandez@gsi.de}

\abst{
Non-destructive Schottky diagnostics at heavy-ion storage rings is broadly exploited to measure revolution frequencies and intensities of stored ions.
The mean Schottky power of incoherent stored ions is expected to be proportional to their number. An empirical examination of this response in few-ion populations is important for interpreting individual occupancy changes. We present an automated staircase reconstruction of the power in a selected frequency band of time-resolved Schottky spectra. The procedure was applied to 7008 selected injections recorded with a resonant cavity pickup at the Experimental Storage Ring at GSI during an 
experiment on the nuclear two-photon decay of the first excited $0^+$ state in bare $^{98}\mathrm{Mo}^{42+}$. The distribution of freely fitted level heights exhibits maxima close to integer multiples of a single-ion power scale up to approximately five ions.
This is consistent with an approximately additive mean power response under the present measurement conditions. The reconstructed occupancy histories also retain population dwell intervals and resolved disappearance times for subsequent lifetime analyses.}

\kword{Schottky spectroscopy, isochronous mass spectrometry, few-ion populations,
change-point detection, nuclear isomers}

\begin{document}
\maketitle

\section{Introduction}
The Experimental Storage Ring (ESR) at GSI Helmholtzzentrum für Schwerionenforschung enables non-destructive monitoring of circulating highly charged ions for extended periods of time~\cite{Litvinov2011, steck2020}. Ions revolve in the ring with about 2\,MHz. At each revolution they excite the Schottky cavity. The noise from the cavity is digitized and Fast Fourier Transformed. The resultant frequency spectra contain harmonics of the ion revolution frequencies~\cite{sanjari2025}. Combining resonant Schottky detection with the isochronous ring operation (S+IMS) allows non-destructive measurements of short-lived species right after injection~\cite{freire2024}. When the first-generation capacitive pickups were used, it was required to average over several seconds to detect electron-cooled single ions~\cite{Litvinov2004,Litvinov2005}. Resonant cavity pickups provide a substantially higher sensitivity and time resolution~\cite{NOLDEN201169,kienle2013,ozturk2019}, so that individual uncooled ions can be followed in the isochronous mode with a resolution of the order of milliseconds~\cite{freire_thesis}. The mean Schottky power of incoherently circulating ions is expected to be proportional to their number~\cite{NOLDEN201169}. For electron-cooled beams, however, the momentum spread was observed to drop abruptly below a few thousand stored ions, and close to this transition the integrated Schottky power was reduced by a factor of two to three, indicating collective effects~\cite{steck1996}. 
Here we present a systematic study of the Schottky power in the few-ion regime recorded by the latest-generation 410\,MHz resonator\,\cite{sanjari2020}. This work complements the analysis of the response of the 245\,MHz detector\,\cite{Buehler}.

The data were recorded at the ESR in experiment G-22-00018 on the nuclear two-photon decay of the $0_2^+$ state in fully ionized $^{98}\mathrm{Mo}^{42+}$~\cite{G2200018}, following the measurement of the isolated two-photon decay in bare $^{72}$Ge~\cite{freire2024}. The $0_2^+\rightarrow0_1^+$ transition cannot proceed through emission of a single real photon. Full ionization suppresses ordinary internal conversion, and its excitation energy of \SI{734.75}{keV} lies below the \SI{1.022}{MeV} electron--positron pair-creation threshold. Two-photon emission is therefore the leading allowed electromagnetic de-excitation channel~\cite{freire2024}. A decay shifts the ion revolution frequency out of the isomer band, producing a decrease in its population-dependent signal. Here, we present an automated procedure that reconstructs single-ion occupancy changes from time-resolved Schottky spectra by representing the band power as a piecewise-constant staircase.

Lifetime analyses based on summing spectrograms from multiple injections
have proved effective~\cite{freire2024}. Resolving individual decay times
and the number of surviving ions provides a complementary route based on
conventional non-destructive decay counting\,\cite{kienle2013,ozturk2019}. Comparing these approaches
can help assess the statistical uncertainties of fits to the summed
signal.

\section{Measurement and staircase reconstruction}
\label{sec:staircase}

\begin{figure}
    \centering
    \includegraphics[width=0.8\linewidth]{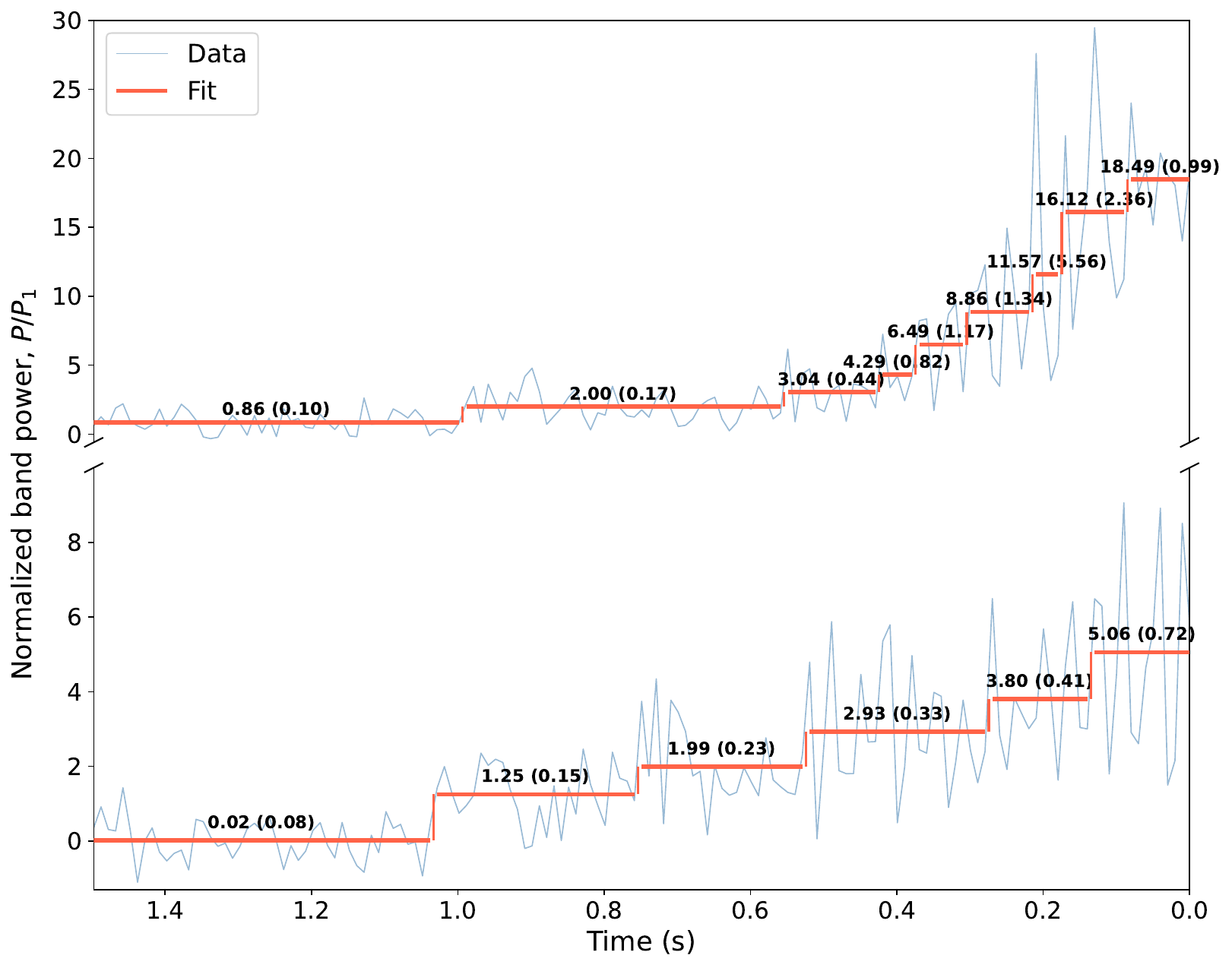}
    \caption{Band power of two example injections in blue and the reconstructed
piecewise-constant staircases in red. Each step corresponds to a change of
the ion number in the frequency band of the $0_2^+$ state. The time axis is reversed due to computational benefits in the staircase fit construction.}
    \label{fig:staircase}
\end{figure}

The analysis uses 7008 selected injections recorded with a \SI{410}{MHz} resonant cavity detector~\cite{sanjari2020}. Each recording is represented by a time-resolved power spectrum, from which the selected isomer-band power is extracted. The ion number discussed below refers to this frequency band.

Let $P_m$, $m=1,\ldots,M$, denote the band-power samples of one injection. Between resolved changes, the mean band power is constant, therefore it can be approximated by a piecewise-constant staircase. The trace is processed in reversed time, starting from the typically longer, less noisy terminal levels. Occupancy is assumed to be non-increasing in forward time, provided that feeding into the selected state frequency band is negligible. All thresholds are expressed in units of the noise $\sigma_0$ of the band power at the end of the recording, where the band is empty in most injections. 

\begin{figure}
    \centering
    \includegraphics[width=0.98\linewidth]{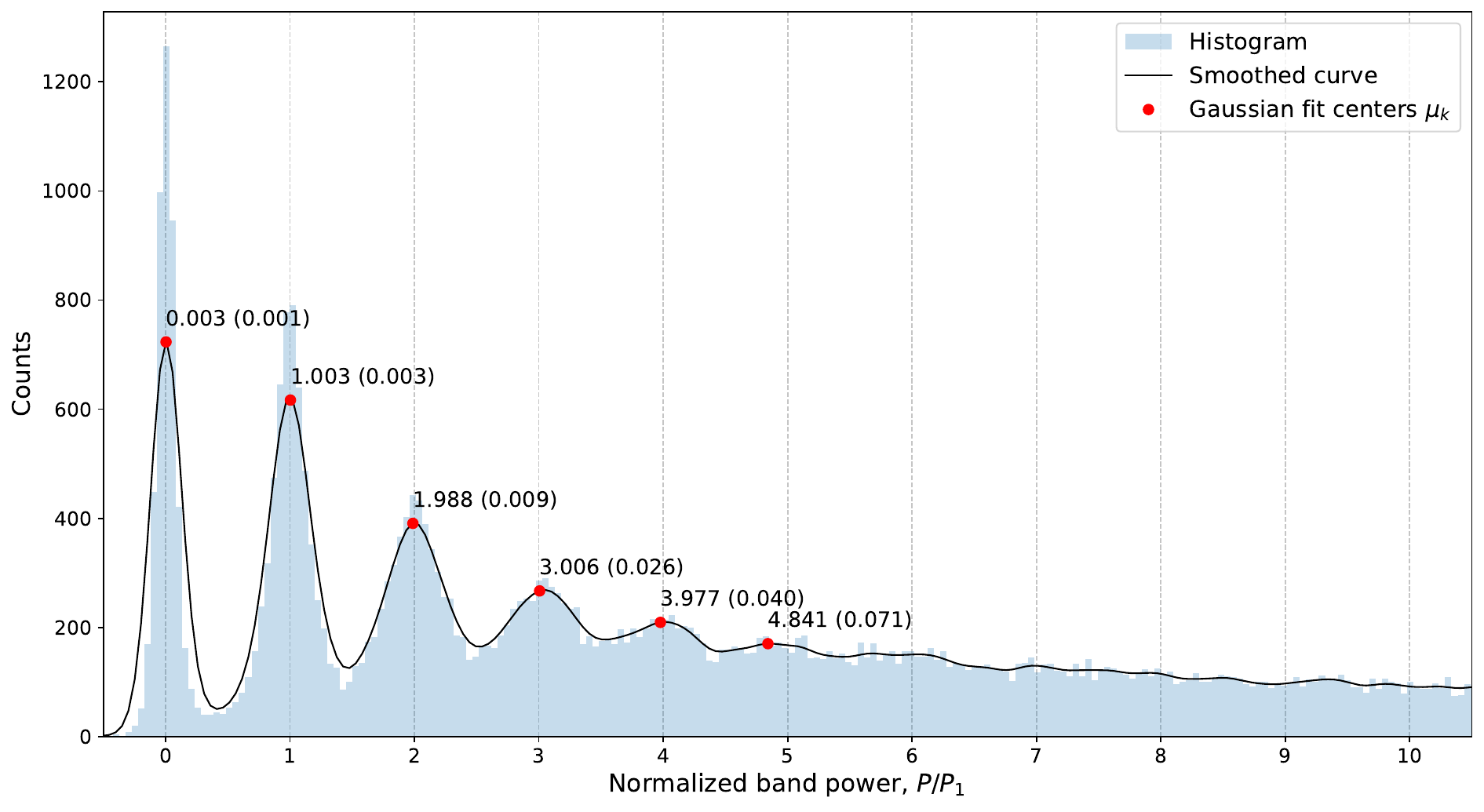}
    \caption{Distribution of the reconstructed level heights $h_k$ in units of
    the single-ion power $P_1$. The peak at zero corresponds to an empty
    frequency band, the further maxima lie at approximately integer multiples of
    $P_1$ corresponding to the number of ions in the band. The statistical uncertainties of the peak positions are obtained from a joint fit of Gaussians to all overlapping peaks. The single-ion power, $P_1 = (1.814 \pm 0.004)\,\sigma_0$, is obtained from a weighted linear fit $\mu_k = \mu_0 + k P_1$ to the fitted peak positions $\mu_k$.

}
    \label{fig:hist}
\end{figure}

For a fixed number of levels $K$, the change points are determined by
minimizing
\begin{equation}
  C_K = \sum_{k=1}^{K} \Big[\, n_k \ln\!\left(\hat\sigma_k^2 + \sigma_0^2\right)
        + \lambda \max\!\left(0,\, n_\mathrm{s}-n_k\right)^2 \Big] ,
  \label{eq:cost}
\end{equation}
where $n_k$ and $\hat\sigma_k^2$ are the sample count and variance in segment $k$. The first term is the negative log-likelihood of Gaussian noise with a separate mean and variance for each segment \cite{truong2020}, which accounts for the increase of the noise with the number of stored ions. $\sigma_0^2$ is used to regularizes nearly flat segments. The second term penalizes short segments. 
Candidate segmentations are obtained using dynamic programming~\cite{jackson2005}, with cuts on segment duration, step height, and the allowed baseline. The number of segments is selected using $C_K+2\alpha(K-1)\ln M$. Long segments are then examined for missed changes
using a two-sample $t$-test, and steps inconsistent with the assumed occupancy evolution are merged. The segment heights $h_k$ are the mean band powers of the resulting segments.

For occupancy reconstruction, each level is assigned an integer ion number $N_k$, using the single-ion power $P_1$ obtained from a weighted linear fit to the peak positions in Fig.\ref{fig:hist}. The assignment minimizes the length-weighted squared deviation of $h_k/P_1$ from $N_k$ under the condition that $N_k$ does not decrease in reversed time. Adjacent levels with the same ion number are merged, so that every remaining change point marks a change of the ion number (Fig.\ref{fig:staircase}).

\section{Few-ion Schottky power levels}
\label{sec:results}
All 7008 selected injections were processed.
Figure~\ref{fig:hist} shows the distribution of the freely fitted level heights $h_k$ in units of the single-ion power $P_1$. Besides a maximum at zero, which corresponds to an empty frequency band, the distribution exhibits distinct maxima at approximately integer multiples of $P_1$ up to five stored ions. This is consistent with the expected linear scaling of the mean Schottky power with ion number~\cite{NOLDEN201169}. The fewer ions are stored, the more accurately the levels are reconstructed, since levels with higher ion numbers become increasingly short and the fluctuations of the band power seem to increase approximately linearly with the ion number. Our results agree qualitatively with the conclusions obtained in the analysis of the 245\,MHz detector\,\cite{Buehler}, in which distinct occupancy peaks emerged only after averaging over \SI{5}{s}. Here, using the 410\,MHz cavity and the staircase reconstruction algorithm, we recover integer-resolved levels up to approximately five uncooled, bare \(^{98}\mathrm{Mo}^{42+}\) ions from segments lasting only a few tens of milliseconds. This shorter analysis timescale is essential for applications to short-lived nuclei.

\section{Conclusion and outlook}
\label{sec:conclusion}
We have presented an automated procedure for reconstructing single-ion
occupancy changes from time-resolved Schottky power spectra. For each injection the band power is fitted by a piecewise-constant staircase, done with the use of dynamic programming, and each level is assigned an
integer ion number. Applied to 7008 injections of fully ionized
$^{98}\mathrm{Mo}^{42+}$ recorded at the Experimental Storage Ring, the reconstructed level heights
show distinct maxima at approximately integer multiples of the single-ion
power up to five stored ions, consistent with the expected linear scaling of
the mean Schottky power with ion number. 
Together with the ion numbers, the dwell times provide the information required for event-resolved lifetime analyses.

Decays that occur in rapid succession cannot always be resolved and are merged into 
a single step. The signal-to-noise ratio is decisive for the quality of the
reconstruction, and isomers with longer half-lives can be reconstructed more
reliably. An application to bare $^{72}\mathrm{Ge}^{32+}$, whose first excited state has a half-life of \SI{23.9(6)}{ms}\cite{freire2024}, was not yet
successful, as the signal-to-noise ratio is lower in this case, due to the lower charge. Further
optimization of the procedure is required to resolve such short-lived and lower signal-to-noise ratio.

\section*{Acknowledgements}
{The methods presented here are being prepared for the analysis of 
the experiment G-22-00018, which was performed at the ESR
storage ring of the GSI Helmholtzzentrum f{\"u}r
Schwerionenforschung, Darmstadt (Germany) in the context of FAIR Phase-0 research program.}

\end{document}